\documentclass[aps,prb,twocolumn,superscriptaddress,showpacs]{revtex4-1}

\usepackage{graphicx}

\begin{document}

\title{Proximity to Fermi-surface topological change in superconducting LaO$_{0.54}$F$_{0.46}$BiS$_2$}

\author{Kensei Terashima}
\affiliation{Graduate School of Natural Science and Technology and Research Laboratory for Surface Science, Okayama University, Okayama 700-8530, Japan}
\affiliation{Research Center of New Functional Materials for Energy Production, Storage, and Transport, Okayama University, Okayama 700-8530, Japan}

\author{Junki Sonoyama}
\affiliation{Graduate School of Natural Science and Technology and Research Laboratory for Surface Science, Okayama University, Okayama 700-8530, Japan}

\author{Takanori Wakita}
\affiliation{Graduate School of Natural Science and Technology and Research Laboratory for Surface Science, Okayama University, Okayama 700-8530, Japan}
\affiliation{Research Center of New Functional Materials for Energy Production, Storage, and Transport, Okayama University, Okayama 700-8530, Japan}

\author{Masanori Sunagawa}
\affiliation{Graduate School of Natural Science and Technology and Research Laboratory for Surface Science, Okayama University, Okayama 700-8530, Japan}

\author{Kanta Ono}
\author{Hiroshi Kumigashira}
\affiliation{High Energy Accelerator Research Organization (KEK), Photon Factory, Tsukuba, Ibaraki, 305-0801, Japan}

\author{Takayuki Muro}
\affiliation{Japan Synchrotron Radiation Research Institute (JASRI)/SPring-8, Sayo, Hyogo 679-5198 Japan}

\author{Masanori Nagao}
\author{Satoshi Watauchi}
\author{Isao Tanaka}
\affiliation{Center for Crystal Science and Technology, University of Yamanashi, Kofu, Yamanashi 400-8511, Japan}

\author{Hiroyuki Okazaki}
\altaffiliation[Present address: ]{WPI-Advanced Institute for Materials Research, Tohoku University, Sendai, 980-8577, Japan}
\author{Yoshihiko Takano}
\affiliation{National Institute for Materials Science, Tsukuba, Ibaraki 305-0047, Japan}

\author{Osuke Miura}
\author{Yoshikazu Mizuguchi}
\affiliation{Department of Electrical and Electronic Engineering, Tokyo Metropolitan University, Hachioji, Tokyo 192-0397, Japan}

\author{Hidetomo Usui}
\author{Katsuhiro Suzuki}
\author{Kazuhiko Kuroki}
\affiliation{Department of Physics, Osaka University, Toyonaka, Osaka 560-0043, Japan}

\author{Yuji Muraoka}
\author{Takayoshi Yokoya}
\affiliation{Graduate School of Natural Science and Technology and Research Laboratory for Surface Science, Okayama University, Okayama 700-8530, Japan}
\affiliation{Research Center of New Functional Materials for Energy Production, Storage, and Transport, Okayama University, Okayama 700-8530, Japan}

\date{\today}

\begin{abstract}  The electronic structure of nearly optimally-doped novel superconductor LaO$_{1-x}$F$_x$BiS$_2$ ({\it x} = 0.46) was investigated using angle-resolved photoemission spectroscopy (ARPES).  We clearly observed band dispersions from 2 to 6 eV binding energy and near the Fermi level (${\it E}_{\rm F}$), which are well reproduced by first principles calculations when the spin-orbit coupling is taken into account.  The ARPES intensity map near ${\it E}_{\rm F}$ shows a square-like distribution around the $\Gamma$(Z) point in addition to electronlike Fermi surface (FS) sheets around the X(R) point, indicating that FS of LaO$_{0.54}$F$_{0.46}$BiS$_2$ is in close proximity to the theoretically-predicted topological change.

\end{abstract}


\maketitle
Low-dimensionality in layered compounds provides a rich soil for novel physical properties and possibility for a variety of ordered states.  Well-known examples are cuprate and iron-pnictide superconductors, where the relationship between the superconductivity and charge/spin degrees of freedom in the superconducting planes are still in hot debate.\cite{1,2}  Recent discovery of a series of Bi-dichalcogenide layered superconductors,\cite{3,4,5,6,7,8,9,10,11,12} Bi$_4$O$_4$(S,Se)$_3$ and (Ln,Sr)(O,F)Bi(S,Se)$_2$ where Ln denotes lanthanoid atoms, has provided a new field of two-dimensional superconductivity with manifested spin-orbital coupling.\cite{13,14}  Among them, LaO$_{1-x}$F$_x$BiS$_2$ exhibits the maximum superconducting transition temperature (${\it T}_{\rm c}$) of 10.6 K at {\it x} $\sim 0.5$ upon pressure.\cite{4}  It consists of slabs of BiS$_2$-La$_2$(O,F)$_2$-BiS$_2$, where the two adjacent BiS$_2$ layers are weakly connected with van der Waals interaction.\cite{14}  The parent compound shows an insulating behavior, while a partial substitution of O with F induces metallic/semiconducting characters in the system and the superconductivity appears.\cite{4}

	Regarding the electronic states of LaO$_{1-x}$F$_x$BiS$_2$, theoretical studies\cite{15,16,17,18,19,20,21} have predicted that the parent compound is a band insulator, and that the carrier is doped into the BiS$_2$ planes upon substitution, resulting in a finite conductivity by quasi one-dimensional electronic structures with nested Fermi surfaces (FSs) of Bi 6{\it p} – S 3{\it p} hybridized bands at {\it x} $\sim 0.5$.  Band structure calculations also predicted presence of a van Hove singularity (vHs) around ($\pi$/2, $\pi$/2) in the Brillouin zone that is located close to Fermi energy (${\it E}_{\rm F}$) at {\it x} $\sim 0.5$, which reminds us of layered systems such as NbSe$_2$ and TaSe$_2$ where the presence of high joint density of states in momentum space resulting from vHs has been argued to be an origin of charge-density-wave.\cite{22,23,24}  In these systems, the relationship between the charge-density-wave gap and the superconducting gap, and the similarity to pseudogap in cuprates have been intensively discussed.\cite{23,24}  Also in Bi-dichalcogenite layered superconductors, theoretical studies point out that such an electronic structure of BiS$_2$ plane is sensitive to interactions of electrons with phonons\cite{19,20,21} and with spin excitations,\cite{15,25,26} which may cause a novel superconductivity in BiS$_2$ layered compounds.  It is expected that the doping-induced ${\it E}_{\rm F}$-crossing of vHs leads to a topological change of FS, namely Lifshitz transition, from electronlike FSs centered at the X(R) point to holelike FSs centered at the $\Gamma$(Z) point, whose signature is reported from macroscopic measurement.\cite{5}  On the other hand, there is a lack of microscopic experimental proof of such characteristic electronic states in BiS$_2$-based compounds.  Recent angle-resolved photoemission spectroscopy (ARPES) of Nd(O,F)BiS$_2$, one of Ln(O,F)BiS$_2$ compounds, reported small electron pockets with no sign of nested FS.\cite{27,28}  The estimated carrier density ({\it x} $\sim 0.1$) was lower than the optimal doping value of {\it x} $\sim 0.5$, thus the electronic structure in the optimal-doping range has remained unknown.

	In this paper, we report the electronic structure of LaO$_{1-x}$F$_x$BiS$_2$ with nearly optimal doping ({\it x} = 0.46) measured by ARPES.  The observed valence bands as well as the band dispersions near ${\it E}_{\rm F}$ show a good agreement with the results of first principles band calculations with spin-orbit coupling, indicating that the electron correlation is moderately low and that the spin-orbit coupling is significantly influencing on the electronic structure where superconductivity occurs.  We have observed a square-like distribution around the $\Gamma$(Z) point in ARPES intensity map near ${\it E}_{\rm F}$ as well as the electron pockets centered at the X(R) point, which is in line with the predicted band dispersions for {\it x} = 0.5 where a vHs exists near ($\pi$/2, $\pi$/2) in the vicinity of ${\it E}_{\rm F}$.  

	LaO$_{1-x}$F$_x$BiS$_2$ single crystals were grown using a CsCl/KCl flux method.  The resistivity\cite{29} showed a sharp drop at ${\it T}_{\rm c}$ , and it became zero at {\it T} (hereafter we call it ${\it T}_{\rm c}^{zero}$) = 3.1 K.  The ratio of the components was estimated by electron probe microanalysis.\cite{29}  ARPES measurements were performed at BL-28A in Photon Factory and BL25SU in SPring-8 with VG scienta electron analyzers, and the energy of a circularly polarized light was set at 70 eV and 880 eV, respectively.  The energy resolutions were set to 30 (200) meV for {\it h}$\nu$ = 70 (880) eV, and the angular resolution was set to 0.2 degrees.  The binding energies of samples were deduced by referencing ${\it E}_{\rm F}$ of gold electronically contacted with samples.  Samples were cleaved and measured in situ under a vacuum better than 3 x 10$^{-8}$ Pa at {\it T}  = 10 K, which was above the transition temperature of the sample.  The samples were stable in the vacuum and showed no sign of degradation during the measurement period of 72h.

	First principles calculations for the electronic structure of {\it x} = 0.5 were performed using Vienna Ab-Initio Simulation Package\cite{30} with the exchange correlation energy functional where the spin-orbit interaction is taken into account,\cite{31} and virtual crystal approximation.\cite{15}  As input parameters, we used the lattice constants of {\it x} = 0.5 reported in the literature.\cite{4}  In the near ${\it E}_{\rm F}$ region (FS of Fig. 1 and Fig. 3, and band dispersion of Fig. 3), we used results obtained from a tight binding model,\cite{15} which was constructed from the first principles band calculation exploiting the maximally localized Wannier orbitals.\cite{32} The Wannier90 code was used for generating the Wannier orbitals using wannier90 package.\cite{33} The model consisted of 24 orbitals for each spin: three {\it p} orbitals in two Bi atoms, three {\it p} orbitals in four S atoms, and three {\it p} orbitals in two O atoms with a frozen window of -2.9 to 0.2 eV with respect to ${\it E}_{\rm F}$, and it well reproduced original first principles calculation results in the energy region shown in the figure.  The model was used to obtain detailed band dispersion with increased {\it k} steps (namely 128$\times$128 points in the $\Gamma$XM plane of the Brillouin zone), which enabled us to compare the calculated results with experimental results taken in the intermediate {\it k}-region.  Calculated FS and band structures for {\it x} = 0.46 were deduced by shifting the chemical potential rigidly by 0.032 eV from the energy position of {\it x} = 0.5 so that the number of occupied electrons become the same as that of doped carriers.

\begin{figure}
\includegraphics{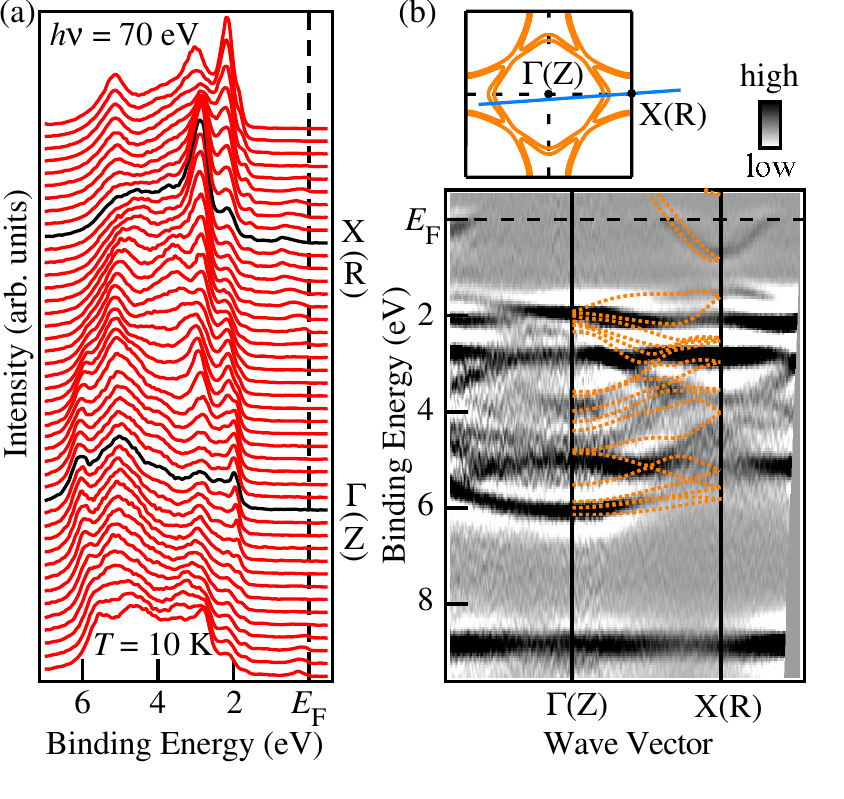}%
\caption{(Color online) Valence band energy distribution curves (a) of LaO$_{0.54}$F$_{0.46}$BiS$_2$ taken along approximately the $\Gamma$(Z)-X(R) direction using photon energy of 70 eV, and their negative values of second derivatives (lower panel of (b)) as a function of binding energy and wave vector. Results of the band calculation for {\it x} = 0.46 along the $\Gamma$-X direction is also plotted in lower panel of (b) as dashed orange lines.  Upper panel of (b) shows a Brillouin zone of LaO$_{1-x}$F$_x$BiS$_2$ and calculated FS for {\it x} = 0.46 (orange lines).  Blue line corresponds to the measured momentum region.
}
\end{figure}

	In Figs. 1(a) and 1(b), we show the valence band energy distribution curves (EDCs) of LaO$_{0.54}$F$_{0.46}$BiS$_2$ measured along the $\Gamma$(Z)-X(R) direction at {\it T} = 10 K and the negative intensity map of their second derivatives as a function of wave vector and binding energy, respectively.  As shown in Figs. 1(a) and 1(b), a number of dispersive bands are observed in the binding energy range of 2 to 6 eV, that are attributed mainly to the O and S {\it p}-dominant bands.\cite{15}  We also observe a nondispersive component at $\sim$ 9 eV, which is a shallow core level of F 2{\it p}.  Along the $\Gamma$(Z)-X(R) direction, the bottom of the valence band is at the $\Gamma$(Z) point, while the top is near the X(R) point with an energy gap of about 0.8 eV to the bottom of the Bi 6{\it p}-dominant band crossing ${\it E}_{\rm F}$.\cite{15,34}  We find a clear discrepancy in the number of electron pockets near the X(R) point between our experimental result and earlier theoretical band dispersions without spin-orbit coupling.  Calculations predict a small electron pocket close to ${\it E}_{\rm F}$ in addition to a large one bottomed at $\sim$ 1 eV, while we do not observe such a small FS in the present experiment.   As pointed out earlier by theoretical studies,\cite{14,21} the effect of spin-orbit coupling in this material is prominent at the X point near ${\it E}_{\rm F}$, where the energy splitting of two downward bands bottomed near the X point are enhanced when the spin-orbit coupling is considered.  In Fig. 1(b), we show our calculated band dispersion along the $\Gamma$X direction (dashed orange lines), which includes the effect of the spin-orbit coupling.  The calculated band dispersion well reproduces the experimental band dispersion (black part of Fig. 1(b)) even quantitatively in such points as (i) the overall width of valence band of $\sim$ 4.5 eV and Bi 6{\it p} band of $\sim$ 0.8 eV (occupied state), (ii) absence of small electron pocket near the X(R) point, and (iii) the band gap of $\sim$0.8 eV between the top of the valence band and the bottom of Bi 6{\it p} band at the X(R) point.  The excellent match between the calculation and the experiment is suggestive of the importance of the spin-orbit coupling in the electronic structure of BiS$_2$ layered compounds.

	From the comparison of the band width between the experiment and the first principles band calculation, the band renormalization factors in both of Bi and S {\it p} bands are close to 1, suggestive of the weak electronic correlations in the system.  We note that similar behavior of band renormalization has been also reported\cite{28} in ARPES study on Nd(O,F)BiS$_2$, implying that BiS$_2$ system with conductive Bi 6{\it p} electrons lies in a weakly correlated regime.

\begin{figure}
\includegraphics{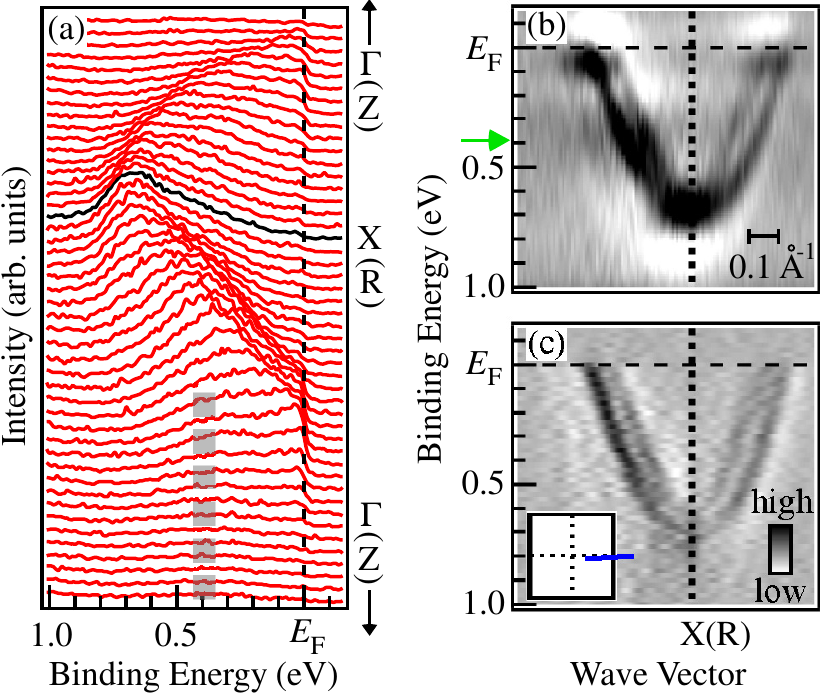}%
\caption{(Color online) EDCs (a) and their second derivatives (b), and second derivatives of MDCs (c) in the vicinity of ${\it E}_{\rm F}$ taken along the $\Gamma$(Z)-X(R) direction near the X(R) point.  Thick line in (a) and green arrow (b) show the energy position of a broad hump.  Inset of (b) shows Brilloiun zone and the measured momentum region (blue line).
}
\end{figure}

	Now we focus on the near-${\it E}_{\rm F}$ region.  Figures 2(a), 2(b), and 2(c) show EDCs taken along the $\Gamma$X$\Gamma$ direction near ${\it E}_{\rm F}$, their second derivatives, and second derivatives of momentum distribution curves (MDCs), respectively.  It is clear from the figures that the downward band bottomed near the X(R) point splits into two branches, which is mainly due to interlayer interactions between adjacent BiS$_2$ layers.\cite{14,27}  As is already mentioned above, we did not observe additional small electron pocket near ${\it E}_{\rm F}$ that appears in the band calculations without the spin-orbit coupling.  Instead, we found an additional component at $\sim$ 400 meV in Figs 2(a) (a thick line) and 2(b) (a green arrow).  On the other hand, the component is not visible in MDC of Fig. 2(c), indicating it's {\it k}-independent (spatially-localized) nature.  The dispersive bands that cross ${\it E}_{\rm F}$ correspond well to the band structure calculations if spin-orbit coupling is considered, implying that this itinerant state is of bulk component.  Next we show further the momentum dependence of both of this Bi 6{\it p}-derived band and the localized band, and discuss the topology of FS.

\begin{figure*}
\includegraphics{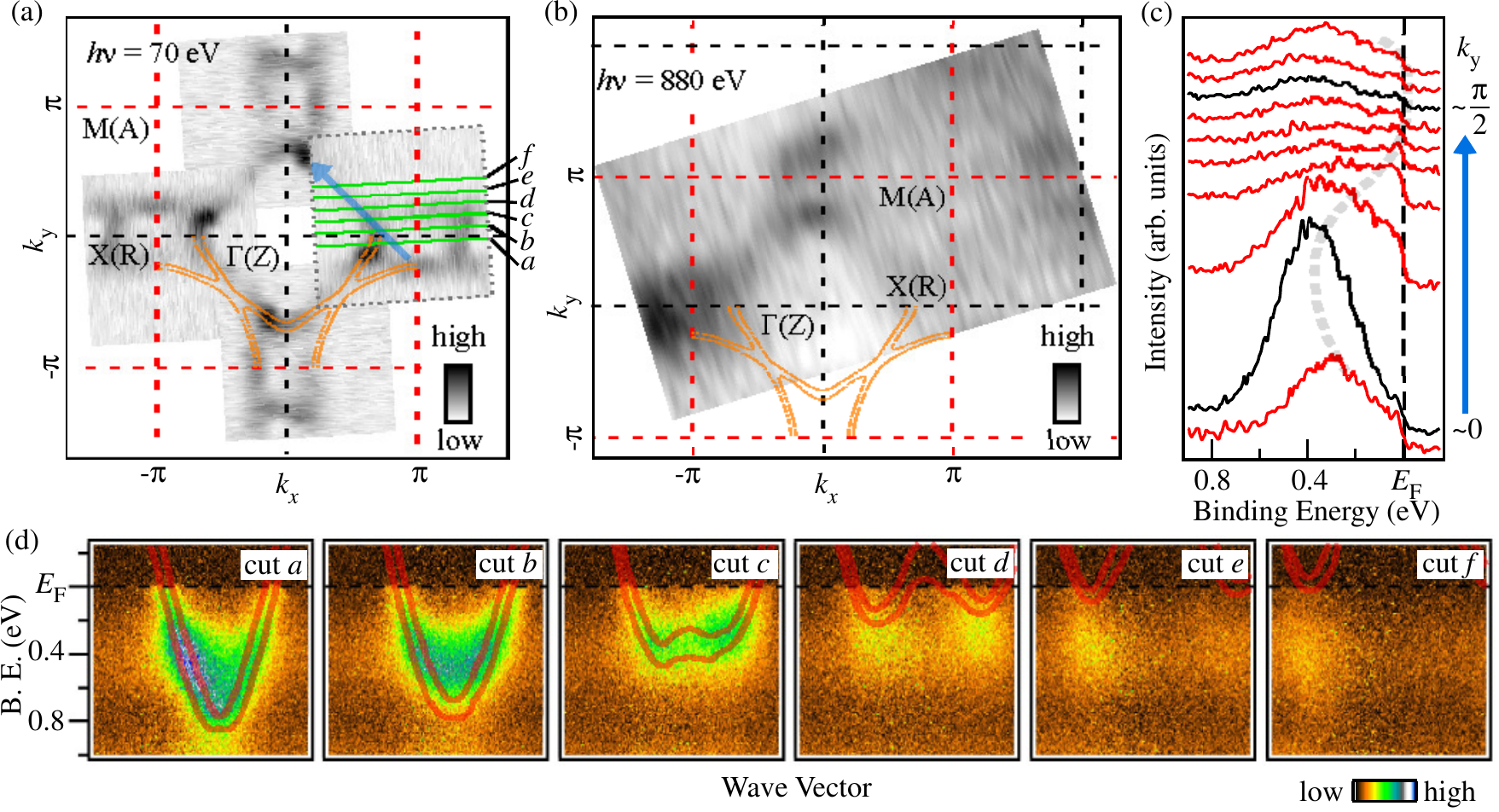}%
\caption{(Color online) (a) ARPES intensity plot integrated over $\pm$40 meV with respect to ${\it E}_{\rm F}$ as a function of two dimensional wave vectors taken at {\it h}$\nu$ = 70 eV, which is symmetrized according to the 4-fold symmetry of the material.  The original data is surrounded by gray dots.  Red dashed lines denote Brillouin zone boundaries, and orange lines show calculated FS for {\it x} = 0.46.  Blue line shows the momentum cut of (c), while green lines show the measured momentum areas of (d).  (b) the same as (a) but taken with use of {\it h}$\nu$ = 880 eV light ($\pm$100 meV with respect to ${\it E}_{\rm F}$).  (c) and (d) EDCs and intensity plots as a function of wave vector and binding energy taken in momentum cuts shown in (a).  Half-transparent gray line in (c) is a guide for the eye, and half-transparent red lines in (d) are calculated bands at corresponding momenta.
}
\end{figure*}

	In Fig. 3(a), we show an ARPES intensity map integrated over $\pm$40 meV with respect to ${\it E}_{\rm F}$ as a function of two dimensional wave vectors.  The data is symmetrized according to the 4-fold symmetry of the crystal, and the area of the original data is surrounded by gray dotted lines in the figure.  The calculated FS for {\it x} = 0.46 is superimposed as orange lines in Figs. 3(a) and 3(b).  We have observed intensity distribution of square-like shape surrounding the $\Gamma$(Z) point as well as the electron pockets centered at the X(R) point.  We also plot a bulk sensitive ARPES intensity map taken with {\it h}$\nu$ = 880 eV in Fig. 3(b), where the integrated energy range was $\pm$100 meV with respect to ${\it E}_{\rm F}$.  The overall intensity distribution in Fig. 3(a) shows close similarity with that of Fig. 3(b) although the energy- and momentum-resolutions are different, namely 30 meV and 0.015 $\AA^{-1}$ for 70 eV-data, while 200 meV and 0.05 $\AA^{-1}$ for {\it h}$\nu$ = 880 eV-data.  Therefore, we concluded that in the near-${\it E}_{\rm F}$ region, the spectra taken with 70 eV light consists mostly of bulk component.  In order to discuss the electronic structures in more detail, we show in Fig. 3(c) a number of representative EDCs along the momentum cut shown as a blue line in Fig. 3(a) where square-like intensity was observed at ${\it E}_{\rm F}$. We also show in Fig. 3(d) ARPES intensity plots as a function of binding energy and momentum taken along cuts shown in Fig. 3(a) as green lines.  In Figs. 3(c) and 3(d), the spectral intensity in different cuts are normalized by photon flux and shown in the same color-scale.  The calculated band dispersions at corresponding momenta are also shown in Fig. 3(d) as half-transparent red lines.  In Fig. 3(d), the overall feature of spectral intensity near ${\it E}_{\rm F}$ is in accordance with calculated bands at a glance, while we recognize two additional structures in higher energies ($\sim$ 0.4 eV) which are not predicted by band calculations.  One is a localized state at $\sim$ 0.4 eV observed in every momentum cuts (also seen in Fig. 2).  The energy scale of the state and it's spatially-localized nature correspond well to the localized state observed in scanning tunneling microscopy on Nd(O,F)BiS$_2$ (Ref. 35).  A surface component is also observed in core level spectra of Bi 4{\it f} (supplemental material), thus we identify the origin of the {\it k-independent} state at $\sim$ 0.4 eV as a surface state.   The other is {\it k-dependent} higher energy component, which is obvious in cuts {\it e} and {\it f}.  It tends to appear in the momentum region where calculated band exists in the vicinity of ${\it E}_{\rm F}$, whose behavior shows close resemblance to an incoherent part of quasiparticles in other compounds\cite{36,37} although at present we are not able to conclude whether it is of bulk nature or not.  It is noted that similar behavior of incoherent part of electrons has been observed in ARPES study on underdoped Nd(O,F)BiS$_2$, whose origin was argued to be polaronic interactions.\cite{27}

	Despite of existence of such higher energy components, the momentum dependence of intensity distribution in Figs. 3(a) and 3(b) as well as the energy dispersion in the vicinity of ${\it E}_{\rm F}$ in Fig. 3(d) show good correspondence with calculated bands.  In the band calculations, Bi 6{\it p}-derived band is predicted to be dispersing downward along the $\Gamma$(Z)-M(A) direction bottomed near ($\pi$/2, $\pi$/2), while the dispersion is upward along the X(R) to the nearest X(R) direction, forming a vHs near ($\pi$/2, $\pi$/2).  When the vHs in Bi 6{\it p} band crosses ${\it E}_{\rm F}$ by electron doping, the topology of FS is predicted to change from electronlike centered at the X(R) point to holelike centered at the $\Gamma$(Z) point.\cite{15}  In the present experiment, apparent upward dispersion has been observed along the X(R) to the nearest X(R) direction in Fig. 3(c) (see the gray guideline).  In Figs. 3(c) and 3(d), the spectral intensity in the vicinity of ${\it E}_{\rm F}$, that is distinguishable from component at higher binding energy by a signature of dip and/or the change of the slope at $\sim$ 0.1 eV, becomes weak as the measured momentum gets close to ($\pi$/2, $\pi$/2).  This behavior of spectral intensity suggests that the vHs is not in occupied side at the present sample.  The finite intensity near ($\pi$/2, $\pi$/2) would correspond to the tail of the high density of states just above ${\it E}_{\rm F}$.  Thus, our ARPES study on nearly optimally-doped La(O,F)BiS$_2$ experimentally proves that a vHs is in proximity to ${\it E}_{\rm F}$, and that the FS of present sample is on the verge of topological change to FS which possesses a good nesting condition along diagonal directions.  We stress that the observed Bi-6{\it p} derived bands are weakly correlated, yet markedly influenced by spin-orbit coupling.  It is expected that our experimental report on these characteristics in Bi 6{\it p}-derived bands serves as a good starting point considering the mechanism for the superconductivity in layered BiS$_2$-based materials.\cite{15,19,20,25,26}

	Finally we compare the present ARPES result with those of BiS$_2$ compounds with other blocking layers.  Earlier ARPES studies on Nd(O,F)BiS$_2$ estimate lower carrier numbers in Bi 6{\it p} bands than the nominal value of {\it x} (Refs. 27,28), whose reason is argued to be due to Bi deficiency.\cite{28}  It is also possible that the valence number of rare earth elements with {\it f} electrons in the blocking layer changes as a function of F concentration as reported in XAS measurement of Ce(O,F)BiS$_2$ (Ref. 38). On the other hand, the observed FS shape as well as the band dispersion on La(O,F)BiS$_2$ is in good accordance with the predicted bands for nominal {\it x} value, thus La(O,F)BiS$_2$ system is likely to be free from such complications.  We also note the relationship between the electronic states and ${\it T}_{\rm c}$s in these compounds.  According to a study of doping dependence on resistivity,\cite{29} the maximum ${\it T}_{\rm c}^{zero}$ of as-grown single crystal La(O,F)BiS$_2$ is $\sim$ 3 K at {\it x} $\sim$ 0.5 to our knowledge.  The present ARPES result is consistent with a scenario where the emergence of ${\it T}_{\rm c}$ is related to the evolution of density of states and/or the nesting condition of FS\cite{19} as long as the Ln atom is fixed to be La.  On the other hand, ARPES studies on Nd(O,F)BiS$_2$ with ${\it T}_{\rm c}^{zero}$ $\sim$ 4 K reported low number of doped electrons in Bi 6{\it p}-derived bands and poor nesting condition of FS.\cite{27,28}  It is expected that there are additional factors that enhance ${\it T}_{\rm c}$ in BiS$_2$-systems other than the nesting condition, such as chemical pressure effect and dimensionality of BiS$_2$ plane controlled by {\it c/a} ratio.\cite{4,6}  In order to fully understand the mechanism of superconductivity in BiS$_2$ systems, further detailed study of electronic states both on La(O,F)BiS$_2$ and Nd(O,F)BiS$_2$ is necessary.

	In summary, we have studied the electronic states of LaO$_{1-x}$F$_x$BiS$_2$ ({\it x} = 0.46).  The observed valence band structure turned out to be well reproduced by first principles band calculation with spin-orbit coupling, suggesting relatively weak electron correlations and marked influence of spin-orbit coupling on BiS$_2$ planes.  We have observed a square-like intensity distribution centered at the $\Gamma$(Z) point at ${\it E}_{\rm F}$ as predicted by theoretical calculations, indicating that the FS is very proximity to topological change and a vHs is in the vicinity of ${\it E}_{\rm F}$ at optimally-doped LaO$_{1-x}$F$_x$BiS$_2$.

\begin{acknowledgments}
ARPES experiments at Photon Factory and SPring-8 were performed under the proposal numbers 2013G703 and 2013A1324, respectively.  This work was partially supported by a Grant-in-Aid for Young Scientists (B) (No, 25800205) from the Ministry of Education, Culture, Sports, Science and Technology of Japan (MEXT).  This work was also partially supported by the Program for Promoting the Enhancement of Research University from MEXT.
\end{acknowledgments}

\begin{widetext}
\newpage
{\bf Supplemental material: Core level spectra of Bi 4{\it f}}\\

	Figures S1(a) and S1(b) show the geometry of photoemission experiment and the emission angle dependence of the Bi 4{\it f} core level photoelectron spectra as a function of binding energy, respectively.  S 2{\it p} peaks are also observed at about 161.5 eV and 162.5 eV.  Both of Bi 4{\it f}7/2 and 4{\it f}5/2 spectra show two peak structures that change their relative intensity when we detect electrons from different angles.  Such an emission angle dependence of spectrum is characteristic to samples which contain surface electronic structures different from the bulk.  We identify that the peak at lower binding energy (~158 eV for Bi 4{\it f}7/2) mainly consists of bulk component and the peak at higher binding energy (~159 eV for Bi 4{\it f}7/2) mainly consists of surface component, since it is expected that the more the emission angle is, the more the photoelectron spectrum becomes surface sensitive.  Although a band structure calculation points out that two adjacent BiS$_2$ layers are mainly connected via van der Waals interaction, our core level spectra imply that the electronic states of the topmost BiS$_2$ layer may not be exactly the same as those of bulk after cleaving.  At present, we have not observed a clear signature of energy separation nor the emission angle dependence in S 2{\it p} peaks despite of the existence of two non-equivalent S sites in a unit cell.

\begin{figure}[b,width=3in]
\includegraphics{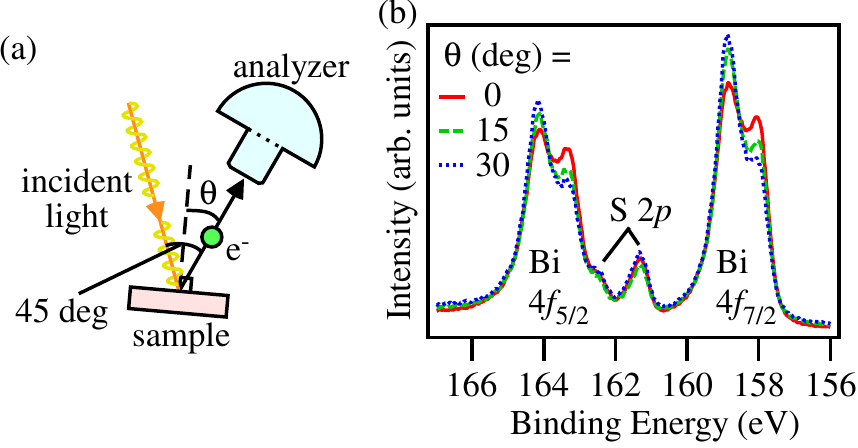}%
\caption{Schematic view of experimental setting both for SPring-8 BL25SU and Photon Factory BL-28A, and definition of emission angle $\theta$.  The angle between incident light and electron analyzer was kept to be 45 degrees.  Emission angle is the angle between the normal direction of sample surface and emitted photoelectrons.  (b) Emission angle dependence of Bi 4{\it f} and S 2{\it p} core level spectrum of LaO$_{0.54}$F$_{0.46}$BiS$_2$.
}
\end{figure}

\end{widetext}
\end{document}